\documentclass[twocolumn,showpacs,nofootinbib]{revtex4}
\usepackage{amssymb}

\usepackage{graphicx}
\usepackage{dcolumn}
\usepackage{bm}
\usepackage{color} 

\begin{document}

\title{Modified-gravity wormholes without exotic matter}
\author{Tiberiu Harko$^1$}
\email{t.harko@ucl.ac.uk}
\author{Francisco S.~N.~Lobo$^{2}$}
\email{flobo@cii.fc.ul.pt}
\author{M. K. Mak$^{3}$}
\email{mkmak@vtc.edu.hk}
\author{Sergey V. Sushkov$^{4}$}
\email{sergey_sushkov@mail.ru}
\affiliation{$^1$Department of Mathematics, University College London, Gower Street,
London WC1E 6BT, United Kingdom}
\affiliation{$^2$Centro de Astronomia e Astrof\'{\i}sica da Universidade de Lisboa, Campo
Grande, Edif\'{i}cio C8 1749-016 Lisboa, Portugal}
\affiliation{$^3$Department of Physics and Center for Theoretical and Computational
Physics, The University of Hong Kong, Pok Fu Lam Road, Hong Kong}
\affiliation{$^4$Institute of Physics, Kazan Federal University, Kremlevskaya Street 18,
Kazan 420008, Russia}

\date{\LaTeX-ed \today}

\begin{abstract}

A fundamental ingredient in wormhole physics is the flaring-out condition at the throat which, in 
classical general relativity, entails the violation of the null energy condition. In this work, we present the 
most general conditions in the context of modified gravity, in which the matter threading the wormhole 
throat satisfies all of the energy conditions, and it is the higher order curvature terms, which may be 
interpreted as a gravitational fluid, that support these nonstandard wormhole geometries. Thus, we explicitly 
show that wormhole geometries can be theoretically constructed without the presence of 
exotic matter, but are sustained in the context of modified gravity.

\end{abstract}

\maketitle






\section{Introduction}

A fundamental property in classical general relativistic wormhole geometries is that these spacetimes are supported by \textit{exotic matter}, which involves a stress-energy tensor $T_{\mu\nu}$ that violates the null energy condition (NEC), i.e., $T_{\mu\nu}k^{\mu}k^{\nu}<0$, where $k^{\mu}$ is \textit{any} null vector \cite{Morris}. More specifically, the NEC imposes that $T_{\mu\nu}k^\mu k^\nu \geq 0$. Indeed, using the theory of embedded hypersurfaces to place restrictions on the Riemann tensor and stress-energy tensor at the throat of the wormhole, Hochberg and Visser demonstrated that
the wormhole throat generically violates the NEC and provided several theorems that generalize the Morris-Thorne results on exotic matter \cite{HochbergVisser1,HochbergVisser2}. In this context, wormhole geometries violate all the pointwise energy conditions and the averaged energy conditions \cite{Visser}. The latter permit localized violations of the energy conditions, as long as on average they hold when integrated along timelike or null geodesics. However, the averaged energy conditions involve a line integral, and therefore do not provide useful information regarding the ``total amount'' of
energy-condition violating matter. This fact prompted the proposal of a ``volume integral quantifier,'' which provides information about the ``total amount'' of energy-condition violating matter in the spacetime \cite{Visser2}. The amount of energy-condition violations is then the extent that these integrals become negative.

Indeed, classical forms of matter are believed to obey the energy conditions, although they are violated by certain quantum fields, such as the Casimir effect and Hawking evaporation \cite{Morris}. Thus, due to its problematic nature, it is useful to minimize the usage of exotic matter, and a wide variety of wormhole solutions have been analyzed in the literature to this effect, ranging from thin-shell wormholes \cite{Visser:1989kg} to rotating \cite{Teo:1998dp} and dynamic wormhole
geometries \cite{HochbergVisser2,dynWH}, and in modified theories of gravity \cite{modgrav1,modgrav2,modgrav3,Lobo:2007qi}. In the latter context, more specifically in $f(R)$ gravity, it was shown that in principle the matter threading the wormholes can be imposed to obey all the energy conditions and it is the higher order curvature terms that are responsible for supporting the geometries \cite{modgrav1}. In braneworlds, the local high-energy bulk effects and the nonlocal corrections from the Weyl curvature in the bulk may induce a NEC violating signature on the brane, while the stress-energy tensor confined on the brane, threading the wormhole, is imposed to satisfy the energy conditions \cite{Lobo:2007qi}. In the curvature-matter coupled generalization of $f(R)$ gravity, exact solutions were found where the nonminimal coupling minimizes the violation of the NEC of matter at the throat \cite{modgrav3}.

It is the purpose of the present paper to generalize the above analysis presenting the most general conditions coming from various modified theories of gravity that are imposed in order to support wormhole geometries. We note that in this context, in principle, one may impose that the matter
stress-energy tensor satisfies the NEC and the respective violations arise from the higher order curvature terms. \\

\section{Wormhole geometries and the energy conditions}

Consider the following wormhole line element in curvature coordinates \cite{Morris}
\begin{equation}
ds^{2}=-e^{2\Phi(r)}dt^{2}+\frac{dr^{2}}{1-b(r)/r}+r^{2}(d\theta ^{2}+\sin
^{2}\theta d\phi ^{2})\,.  \label{WHmetric}
\end{equation}
The redshift function $\Phi(r)$ must be finite everywhere to avoid the
presence of event horizons. In order to have a wormhole geometry, the shape
function $b(r)$ must obey the flaring-out condition of the throat $r_0$, which is
translated by $(b- b^{\prime}r)/b^{2}>0$ \cite{Morris}. At the throat, we have
$b(r_{0})=r_{0}$, and the condition $b^{\prime}(r_{0})<1$ is imposed. Note that the flaring-out condition has a purely geometric nature. However, in
classical general relativity, through the Einstein field equation, one can
deduce that the matter threading the wormhole throat violates the NEC.

Generally, the NEC arises when one refers back to the Raychaudhuri
equation, given by
\begin{equation} \label{Raych}
\frac{d\theta}{d\tau}= - \frac{1}{2}\,\theta^2 -
\sigma_{\mu\nu}\sigma^{\mu\nu} + \omega_{\mu\nu}\omega^{\mu\nu} -
R_{\mu\nu}k^{\mu}k^{\nu} \;,
\end{equation}
where $R_{\mu\nu}$ is the Ricci tensor, and $\theta\,$, $\sigma^{\mu\nu}$
and $\omega_{\mu\nu}$ are, respectively, the expansion, shear and rotation
associated to the congruence defined by the null vector field $k^{\mu}$.
The Raychaudhuri equation is also a purely geometric statement, and as
such it makes no reference to any gravitational field equations.
Now, the shear is a ``spatial'' tensor with
$\sigma^2 \equiv \sigma_{\mu\nu}\sigma^{\mu\nu}\geq 0$ and $\omega_{\mu\nu}
\equiv 0$ for any hypersurface orthogonal congruences, so that the condition for attractive gravity 
reduces to $R_{\mu\nu}k^{\mu}k^{\nu}\geq 0$. The positivity of this latter quantity ensures
that geodesic congruences focus within a finite value of the parameter
labeling points on the geodesics. In general relativity, contracting
both sides of the Einstein field equation $G_{\mu\nu}\equiv
R_{\mu\nu}-\frac12 Rg_{\mu\nu}=\kappa^2 T_{\mu\nu}$ with
any null vector $k^\mu$, one can write the above condition in terms of
the stress-energy tensor given by $T_{\mu\nu}k^\mu k^\nu \ge 0$.

In modified theories of gravity the gravitational field equations can be
rewritten as an effective Einstein equation, given by $G_{\mu\nu}=\kappa^2
T_{\mu\nu}^{\mathrm{eff}}$, where $T_{\mu\nu}^{\mathrm{eff }}$ is
an effective stress-energy tensor containing the matter stress-energy tensor
$T_{\mu\nu}$ and the curvature quantities, arising from the specific modified
theory of gravity considered. Now, the positivity condition
$R_{\mu\nu}k^{\mu}k^{\nu}\geq 0$ in the Raychaudhuri equation provides the
\textit{generalized} NEC, $T^{\mathrm{eff}}_{\mu\nu} k^\mu k^\nu\geq 0$, through
the modified gravitational field equation.

By definition (see Ref. \cite{HochbergVisser1}) the wormhole throat has to defocus a
null geodesic congruence. Therefore, the necessary condition to have a wormhole
geometry is the violation of the generalized NEC, i.e.,
$T^{\mathrm{eff}}_{\mu\nu} k^\mu k^\nu< 0$. In classical general relativity this
simply reduces to the violation of the usual NEC, i.e., $T_{\mu\nu} k^\mu k^\nu<
0$. However, in modified theories of gravity, one may in principle impose that the matter
stress-energy tensor satisfies the standard NEC, $T_{\mu\nu} k^\mu k^\nu\geq 0$,
while the respective generalized NEC is necessarily violated, $T^{\mathrm{eff}}_{\mu\nu}
k^\mu k^\nu< 0$, in order to ensure the flaring-out condition.

Note that instead of a null geodesic congruence we can consider a congruence of
timelike geodesics. In this case the corresponding Raychaudhuri equation reads
\begin{equation}
\frac{d\hat\theta}{d\tau}= - \frac{1}{3}\hat\theta^2 -
\hat\sigma_{\mu\nu}\sigma^{\mu\nu} + \hat\omega_{\mu\nu}\omega^{\mu\nu} -
R_{\mu\nu}u^{\mu}u^{\nu},
\end{equation}
where $\hat\theta$, $\hat\sigma^{\mu\nu}$ and $\hat\omega_{\mu\nu}$ are,
respectively, the expansion, shear and twist of the congruence defined by the
timelike unit vector field $u^{\mu}$ normalized to unit length, $u_\mu
u^\mu=-1$. The positivity condition $R_{\mu\nu}u^{\mu}u^{\nu}\geq 0$
focusses the timelike congruence and ensures an attractive nature of
gravity. In classical general relativity, using Einstein's equation, we can
write this condition as $ T_{\mu\nu}u^\mu u^\nu\geq -\frac12 T$,
where $u^\mu$ is any timelike vector. This assumption is known as the strong
energy condition (SEC). Its violation is a necessary condition to have a
wormhole geometry.  \\

\section{Wormholes in generalized modified gravity}
\label{2}

Consider the generalized gravitational field equations for a large class of modified theories of gravity, given by the following field equation
\begin{equation}
g_1(\Psi^i)(G_{\mu\nu}+H_{\mu\nu})-g_2(\Psi^j)\,T_{\mu\nu}=\kappa^2\,T_{\mu%
\nu}\,,
  \label{generalfieldeq}
\end{equation}
where $H_{\mu\nu}$ is an additional geometric term that includes the geometrical modifications inherent in the modified gravitational theory under consideration; $g_i(\Psi^j)$ ($i=1,2$) are multiplicative factors that modify the geometrical sector of the field equations, and $\Psi^j$ denote generically curvature invariants or gravitational fields such as scalar fields; the term $g_2(\Psi^i)$ covers the coupling of the curvature invariants or the scalar fields with the matter stress-energy tensor, $T_{\mu\nu}$.

It is useful to rewrite this field equation as an effective Einstein field
equation, as mentioned above, with the effective stress-energy tensor, $T_{\mu\nu}^{\mathrm{eff}}$, given by
\begin{equation}
T_{\mu\nu}^{\mathrm{eff}} \equiv \frac{1+\bar{g}_2(\Psi^j)}{g_1(\Psi^i)}
\,T_{\mu\nu} -\bar{H}_{\mu\nu}\,,
\end{equation}
where $\bar{g}_2(\Psi^j)=g_2(\Psi^j)/\kappa^2$ and $\bar{H}
_{\mu\nu}=H_{\mu\nu}/\kappa^2$ are defined for notational convenience.

In modified gravity, the violation of the generalized NEC, $T^{\mathrm{eff}}_{\mu\nu} k^\mu k^\nu < 0$, implies the following restriction
\begin{equation}
\frac{1+\bar{g}_2(\Psi^j)}{g_1(\Psi^i)}\,T_{\mu\nu} k^\mu k^\nu < \bar{H}%
_{\mu\nu}k^\mu k^\nu \,.
\end{equation}
For general relativity, with $g_1(\Psi^j)=1$ , $g_2(\Psi^j)=0$, and $%
H_{\mu\nu}=0$, we recover the standard violation of the NEC for the
matter threading the wormhole, i.e., $T_{\mu\nu} k^\mu k^\nu < 0$.

If the additional condition $[1+\bar{g}_2(\Psi^j)]/g_1(\Psi^i)>0$ is met, then one obtains a general
bound for the normal matter threading the wormhole, in the context of modified theories of gravity, given by
\begin{equation}
0 \leq T_{\mu\nu} k^\mu k^\nu < \frac{g_1(\Psi^i)}{1+\bar{g}_2(\Psi^j)}\,
\bar{H}_{\mu\nu}k^\mu k^\nu \,.
\end{equation}

Analogously, in modified gravity, wormhole solutions also violate the generalized
SEC, i.e., $ T^{\mathrm{eff}}_{\mu\nu}u^\mu
u^\nu< -\frac12 T^{\mathrm{eff}}$, which implies the
following bound on the matter stress-energy tensor
\begin{equation}
\frac{1+\bar{g}_2(\Psi^j)}{g_1(\Psi^i)}\left[T_{\mu\nu} u^\mu u^\nu-\frac12
T\right] < \bar{H}_{\mu\nu}u^\mu u^\nu -\frac12 \bar H.
\end{equation}
Now, one may demand that the latter condition is fulfilled even if
the matter stress-energy tensor satisfies the usual SEC,
$T_{\mu\nu}u^\mu u^\nu-\frac12 T\geq 0$, or the weak energy condition (WEC)
$T_{\mu\nu}u^\mu u^\nu\geq 0$.

In order for normal matter to satisfy the WEC, to have a positive energy density, one also needs to impose the following relationship
\begin{equation}
T_{\mu\nu}u^\mu u^\nu =
\frac{g_1(\Psi^i) }{\kappa^2+g_2(\Psi^j)}
\left(G_{\mu\nu} + H_{\mu\nu} \right) u^\mu u^\nu \geq 0 \,.
\end{equation}
Imposing $T_{\mu\nu}u^\mu u^\nu\geq 0$ entails a restriction on the geometry arising from the modified gravity under consideration. If the normal matter is given by a diagonal stress-energy tensor, i.e., $T^{\mu}{}_{\nu}={\rm diag}[-\rho(r),p_r(r),p_t(r),p_t(r)]$, one can physically interpret $T_{\mu\nu}u^\mu u^\nu$ as the energy density measured by any timelike observer with four-velocity $u^\mu$. This definition is useful as using local Lorentz transformations it is possible to show that $T_{\mu\nu}u^\mu u^\nu \geq 0$ implies that the energy density is positive in all local frames of reference. Thus, the standard WEC imposes that $\rho \geq 0$ and $\rho + p_i \geq 0$ (where $i=r,t$).  \\

\subsection{$f(R)$ gravity}

A well-known modification of general relativity is $f(R)$ gravity with the following action
\begin{equation}
S=\int d^4x\sqrt{-g} \left\{\frac{1}{2\kappa^2}f(R)+{\cal L}_m\right\}.
\end{equation}
The stress-energy tensor of matter is defined as \cite{LaLi}
$T_{\mu \nu }=-\frac{2}{\sqrt{-g}}\frac{\delta \left( \sqrt{-g}\mathcal{L}%
_{m}\right) }{\delta g^{\mu \nu }}$,
and we consider that the matter Lagrangian density $\mathcal{L}_{m}$ only
depends on the metric tensor components $g_{\mu \nu }$, and not on its
derivatives. 

In this case, the field equations are given by Eq. (\ref{generalfieldeq}), with
the relationships $g_{1}(\Psi ^{i}) =f_{R}(R)$, $g_{2}(\Psi ^{j})=0$, and
\begin{equation}
H_{\mu \nu } =\frac{1}{f_{R}}\left[ \frac{1}{2}(Rf_{R}-f) g_{\mu\nu}
 -\nabla _{\mu
}\nabla _{\nu }f_{R}+g_{\mu \nu }\Box f_{R}\right].
\end{equation}
where $f_R=df/dR$.

The generic condition for the violation of the generalized NEC in $f(R)$ gravity is given by
\begin{equation}\label{ineq1}
\frac{1}{f_{R}}T_{\mu \nu }k^{\mu }k^{\nu }<-\frac{1}{\kappa ^{2}f_{R}}%
\,k^{\mu }k^{\nu }\,\nabla _{\mu }\nabla _{\nu }f_{R}\,.
\end{equation}%
It is worth noting that, depending on a particular form of $f(R)$, the latter
inequality could be fulfilled even if $T_{\mu \nu }k^{\mu }k^{\nu }>0$.
In particular, if $f_{R}>0$, then the bound
$0\leq \kappa ^{2} T_{\mu \nu }k^{\mu }k^{\nu }<-k^{\mu }k^{\nu
}\,\nabla _{\mu }\nabla _{\nu }f_{R}$ is imposed.

As a specific example, consider $R^2$ gravity with $f(R)=R+\frac12\alpha
R^2$, so that the second term dominates for strong curvatures as is the case at the wormhole throat and its neighborhood. In this case, Eq. (\ref{ineq1}) reads
\begin{equation}\label{R2}
\frac{T_{\mu \nu}k^{\mu }k^{\nu }}{1+\alpha R}<-\frac{\alpha k^{\mu}k^{\nu}R_{;\mu\nu}}
{\kappa ^{2}(1+\alpha R)}.
\end{equation}%
In particular, in the wormhole metric (\ref{WHmetric}) with $\Phi(r)\equiv0$, so that the curvature scalar is given by $R=2b'/r^2$, then inequality (\ref{R2}) evaluated at the throat $r_0$ takes the form
\begin{equation}\label{th2}
\frac{T_{\mu \nu}k^{\mu }k^{\nu }|_{r_0}}{r_0^2+2\alpha b'_0} < \frac{\alpha
(1-b'_0)(2b'_0-r_0b''_0)}
{\kappa ^{2}r_0^4(r_0^2+2\alpha b'_0)}.
\end{equation}

It is obvious that for $\alpha=0$ we have $T_{\mu \nu }k^{\mu }k^{\nu }|_{r_0}<0$.
However, generally we can choose $\alpha$ so that Eq. (\ref{th2}) is fulfilled
even if $T_{\mu \nu }k^{\mu }k^{\nu }\geq 0$.

\subsection{Curvature-matter coupling}
\label{2}

Now let us consider the modified theory of gravity with an explicit curvature-matter coupling given by the following action \cite{Bertolami:2007gv}
\begin{equation}
S=\int d^4x\sqrt{-g}\left\{\frac{1}{2\kappa^2}f(R)+\big[1+\lambda h(R) \big]
\mathcal{L}_m \right\},
\end{equation}
where $f(R)$ and $h(R)$ are arbitrary functions of the Ricci scalar $R$. The
coupling constant $\lambda$ characterizes the strength of the interaction
between $h(R)$ and the matter Lagrangian.

Taking into account the modified Einstein equation, the effective
stress-energy tensor is given by
\begin{eqnarray}
T_{\mu\nu}^{\mathrm{eff}}&=&\frac{1}{{\cal L}_{\rm c}} \Big\{(1+\lambda h)\,T_{\mu\nu}
+\frac1{2\kappa^2}\left(f-{\cal L}_{\rm c} R
\right)g_{\mu\nu}
\nonumber \\
&&
- \frac{1}{\kappa^2}\left(g_{\mu\nu}\nabla_\alpha \nabla^\alpha-\nabla_\mu \nabla_\nu
\right) {\cal L}_{\rm c} \Big\}
\,,
\end{eqnarray}
where $f_{R}=d f/d R$ and $h_{R}=d h/d R$, and we have defined
${\cal L}_{\rm c} \equiv  f_{R}+2\lambda \kappa^2 h_{R} \mathcal{L}_m$ for notational simplicity.

The general condition to have wormhole geometries, $T_{\mu\nu}^{\mathrm{eff}%
}k^\mu k^\nu<0$, reduces to
\begin{eqnarray}\label{cond2}
\frac{1+\lambda h}{{\cal L}_{\rm c} } T_{\mu\nu}\,k^\mu k^\nu
<-\frac{k^\mu k^\nu\;\nabla_\mu \nabla_\nu \,{\cal L}_{\rm c}}{\kappa^2 {\cal L}_{\rm c}}.
\end{eqnarray}

As a specific example, consider the model with $f(R)=h(R)=R$ and ${\cal
L}_m=-\rho$ \cite{Bertolami:2008ab}, where $\rho$ is the energy density of matter. In this case,
Eq. (\ref{cond2}) yields
\begin{equation}\label{cond22}
\frac{1+\lambda R}{1-2\kappa^2\lambda\rho} T_{\mu\nu}\,k^\mu k^\nu
<\frac{2\lambda k^\mu k^\nu\rho_{;\mu\nu}}{1-2\kappa^2\lambda\rho}.
\end{equation}

For the wormhole geometry (\ref{WHmetric}) with $\Phi(r)\equiv0$,
and choosing the null vector $k^\mu=(1,\sqrt{1-b/r},0,0)$, so that $
k^\mu k^\nu\nabla_\mu\nabla_\nu \rho=[2(r^2-rb)\rho''+(b-rb')\rho']/(2r^2)
$, the inequality (\ref{cond22}) evaluated at the throat $r_0$, is as follows
\begin{equation}\label{cond23}
\frac{r_0^2+2\lambda b'_0}{1-2\kappa^2\lambda\rho_0} \;T_{\mu\nu}\,k^\mu k^\nu |_{r_0}
<\frac{\lambda r_0\rho'_0 (1-b'_0)}{1-2\kappa^2\lambda\rho_0}.
\end{equation}
Again, it is seen that $T_{\mu\nu}\,k^\mu k^\nu |_{r_0}<0$ provided $\lambda=0$.
However, generally one can choose the specific parameters of the model so that
the restriction (\ref{cond23}) could be fulfilled even if $T_{\mu \nu }k^{\mu
}k^{\nu }\geq 0$. \\

\subsection{$F\left(R,\mathcal{L}_{m}\right)$ gravity}

In this section, consider the following action for the generalized modified
gravitational theory \cite{Harko:2010mv}
\begin{equation}\label{action3}
S=\int d^4x\sqrt{-g}\, F\left(R,\mathcal{L}_{m}\right),
\end{equation}
where $F\left(R,\mathcal{L}_{m}\right)$ is an arbitrary function of the Ricci scalar $R$, and of matter Lagrangian density, $\mathcal{L}_{m}$. This theory generalizes the $f(R)$ gravity models and the
curvature-matter couplings \cite{Bertolami:2007gv}.

For $F\left( R,\mathcal{L}_{m}\right) $ gravity, we have the
following relationships $H_{\mu \nu } =\frac{1}{F_{R}}\left[\frac{1}{2}\bar{F} g_{\mu \nu }
+\Box_{\mu\nu} F_{R}\right] $,
$g_{1}(\Psi ^{i}) =F_{R}$ and $g_{2}(\Psi ^{j})=F_{\mathcal{L}_{m}}/2-\kappa ^{2}$,
where we have defined $\bar{F} \equiv (F_{R}R+F_{\mathcal{L}%
_{m}}\mathcal{L}_{m}-F)$,
$\Box_{\mu\nu} \equiv g_{\mu \nu }\nabla _{\alpha }\nabla ^{\alpha }-\nabla _{\mu }\nabla_{\nu }$ and denoted $F_{R}=\partial F/\partial R$ and $F_{\mathcal{L}_{m}}=\partial F /\partial \mathcal{L}_{m}$, respectively.

The gravitational field equation may be written as an effective
Einstein field equation, with $T_{\mu\nu}^{\mathrm{eff}}$ given by
\begin{eqnarray}
T_{\mu\nu}^{\mathrm{eff}}=\frac{1}{2F_R}\big[\bar{F} g_{\mu\nu}
-2\Box_{\mu\nu}\,F_R +F_{\mathcal{L}_{m}}T_{\mu\nu} \big]\,.
\end{eqnarray}
Note that the divergence of the stress-energy tensor $T_{\mu \nu}$ is given by
$\nabla ^{\mu }T_{\mu \nu }=2\nabla ^{\mu }\ln \left( F_{\mathcal{L}_{m}} \right) \frac{\partial \mathcal{L}_{m}}{\partial g^{\mu \nu }}$, which translates an explicit exchange of energy and momentum between the matter and the higher order curvature terms.
The covariant conservation of the stress-energy tensor, $\nabla^{\mu }T_{\mu \nu }=0$, provides an effective functional relation between the matter Lagrangian density and the function $F_{\mathcal{L}_{m}}\left(R,\mathcal{L}_{m}\right)$, given by $\nabla^{\mu }\ln \left( F_{\mathcal{L}_{m}} \right) \;\partial \mathcal{L}_{m}/\partial g^{\mu \nu }=0$.
Thus, in principle, once the matter Lagrangian density is known, by an appropriate choice of the function $F\left( R,\mathcal{L}_{m}\right) $ one can construct conservative models
with arbitrary curvature-geometry couplings \cite{Harko:2010zi}.

Now, contracting $T_{\mu\nu}^{\mathrm{eff}}$ with \textit{any} null vector, $k^\mu$, the essential
condition to support wormhole geometries is $T_{\mu\nu}^{\mathrm{eff}}k^\mu k^\nu<0$, which is 
given by
\begin{eqnarray}
\frac{F_{\mathcal{L}_{m}}}{F_R}T_{\mu\nu} k^\mu k^\nu <
-\frac{2 k^\mu k^\nu \nabla_\mu \nabla_\nu F_R}{F_R}\,.  \label{genNEC}
\end{eqnarray}
For $f(R,\mathcal{L}_{m})=R/2\kappa^2+\mathcal{L}_{m}$, then (\ref{genNEC}) reduces to the standard violation of the NEC in general relativity \cite{Morris}.

Thus, the general condition for wormhole spacetimes is given by inequality (\ref{genNEC}), and in principle one may now construct specific solutions, either by considering a specific form for $F(R,\mathcal{L}_{m})$, and by imposing an equation of state of the matter stress-energy tensor, and/or considering choices for the metric functions $\Phi(r)$ and $b(r)$. We leave this analysis for  a future publication. 

\section{Discussions and final remarks}
\label{6}

Despite the fact that the flaring-out condition, in classical general relativity, through the Einstein field
equation necessarily entails the violation of the NEC, in modified theories of gravity it is the {\it
generalized} NEC that is violated. The latter involves an effective stress-energy tensor that includes the
matter stress-energy tensor and higher curvature terms, which may be interpreted as a gravitational
fluid. Thus, in this work, we have explicitly shown that one may impose that the normal matter
stress-energy tensor satisfies all the standard energy conditions and presented very general restrictions
on the wormhole geometry arising from the modified theory of gravity under consideration. Indeed, this
is translated through general inequalities showing that the higher curvature terms sustain the wormhole
geometries.

In addition to this, we have considered specific cases of modified theories of gravity considered in the literature, namely, $f(R)$ gravity, the curvature-matter coupling and the $F(R,\mathcal{L}_{m})$ generalization. In the first two theories, we analyzed specific cases and showed explicitly that one may choose the parameters of the theory such that the matter threading the wormhole throat satisfies the energy conditions. Thus, this shows that one may theoretically construct wormhole geometries without the use of exotic matter, although it is the higher order curvature terms arising in the modified theories of gravity that sustain these exotic spacetimes.

It would also be interesting to analyze whether the various modified gravitational models required 
for a wormhole have any serious instabilities. For instance in $R+\alpha R^2$ gravity, it was claimed in
\cite{Coule:1993wc} that a Lorentzian wormhole would require $\alpha < 0$, i.e., the existence of the 
sign that corresponds to an unbounded from below potential or tachyonic kinetic term in the 
conformally related scalar field model (see also \cite{Bronnikov:2010tt}). In this context, it would be 
interesting to consider whether a modified gravitational model could effectively violate the null energy 
condition without such a related instability. Although this analysis lies outside the scope of the present 
paper, work along these lines is currently under way. 

\section*{Acknowledgments}

FSNL acknowledges financial support of the Funda\c{c}\~{a}o para a Ci\^{e}%
ncia e Tecnologia through the grants CERN/FP/123615/2011 and
CERN/FP/123618/2011. SVS acknowledges financial support of the Russian Foundation for Basic Research through the grant No. 11-02-01162.

\end{document}